\documentclass[twocolumn,aps,prc,showpacs,nofootinbib,floatfix]{revtex4}
\usepackage{graphicx}
\usepackage{dcolumn}
\usepackage{bm}
\usepackage{longtable}
\begin{document}
\title{Percolation of Color Sources and the Equation of State of QGP
in central Au-Au collisions at $\sqrt{s_{NN}}=$200 GeV}
\author{R. P. Scharenberg, B. K. Srivastava, A. S. Hirsch}
\address{Department of Physics, Purdue University \\
West Lafayette, IN-47907, USA}

\date{\today}
\begin{abstract}
 The Color String Percolation Model (CSPM) is used to determine the equation of state (EOS) of the Quark-Gluon Plasma (QGP) produced in central Au-Au collisions at $\sqrt{s_{NN}}$ = 200 A GeV using STAR data at RHIC. 
When the initial density of interacting colored strings exceeds the 2D percolation threshold a cluster is formed, which defines the onset of  color deconfinement. These interactions also produce fluctuations in the string tension which transforms the Schwinger particle (gluon) production mechanism into a maximum entropy thermal distribution analogous to QCD Hawking-Unruh radiation.
The single string tension is determined by identifying the known value of the universal hadron limiting temperature $T_{c}$ = 167.7 $\pm$ 2.6  MeV with the CSPM temperature at the critical percolation threshold parameter $\xi_{c}$  =1.2. At mid-rapidity the initial Bjorken energy density and the initial temperature determine the number of degrees of freedom consistent with the formation of a $\sim$ 2+1 flavor QGP.  An analytic expression for the equation of state, the sound velocity $C_{s}^{2}(\xi)$ is obtained in CSPM. The CSPM $C_{s}^{2}(\xi)$ and the bulk thermodynamic values energy density $\varepsilon /T^{4}$ and  entropy density $s /T^{3}$ are in excellent agreement in the phase transition region with recent lattice QCD simulations (LQCD) by the HotQCD Collaboration.\\
\end{abstract}
\maketitle  
\newpage

All high energy soft multi-hadron interactions exhibit thermal patterns of 
abundances characterized by the same temperature, independent of the center of mass energy \cite{bec1}. The hadron limiting temperatures were measured by statistical thermal 
analyses that fit the data with a minimum of parameters \cite{bec1}.  
In heavy ion collisions it may be plausible that multiple parton interactions produce a thermalized system. In (p, p), (p, $\bar{p}$) and (p, A) collisions multiple parton interactions are not likely to thermalize the system. In ($e^{+} e^{-}$) annihilation where thermal behavior is observed, the multi-parton mechanism may also be an inappropriate explanation \cite{bec2,satz2,stat1}. 

The determination of the EOS of hot, strongly interacting matter is one of the main challenges of strong interaction physics (HotQCD Collaboration) \cite{bazavov}. Recent LQCD calculations for the bulk thermodynamic observables, e.g. pressure, energy density, entropy density and for the sound velocity have been reported \cite{bazavov}. In the present work a percolation model coupled with hydrodynamics has been utilized to calculate these quantities and compare them with the LQCD results.

The  CSPM describes the initial collision of two heavy ions in terms of a dense pattern of interacting colored strings which start to overlap to form clusters \cite{pajares1,pajares2}. The interactions of strings reduces the hadron multiplicity $\mu$  and increases the average transverse momentum ${\langle  p_{T}^{2} \rangle}$ of these hadrons, so that the total transverse momentum is conserved .
The observables $\mu$, and $\langle p_{t}^{2} \rangle$ are directly related to the field strength in the string and thus to the generating color.  For a cluster of n strings 
 \begin{equation}                                      
 n = \frac{\mu}{\mu_{0}} \frac{\langle p_{t}^{2}\rangle}{\langle p_{t}^{2}\rangle_{1}}    \end{equation} 
where $\mu_{0}$ is  the multiplicity of a single string and $\langle p_{t}^{2}\rangle_{1}$ is the average transverse momentum squared of a single string.
 The CSPM model calculation for hadron multiplicities and momentum spectra was found to be in excellent agreement with experiment \cite{diasde}. Within the framework of clustering of color sources, the elliptic flow, $\it v_{2}$, and the dependence of the nuclear modification factor on the azimuthal angle show reasonable agreement with the RHIC data \cite{bautista}. The critical density of percolation is related to the effective critical temperature and thus percolation may be the way to achieve deconfinement in the heavy ion collisions \cite{pajares3}. An additional important check of this interacting string approach was provided by the measurement of Long Range forward backward multiplicity Correlations (LRC) by the STAR group at RHIC \cite{LRC}.

The two dimensional (2D) percolation threshold identifies the hadron to quark-gluon percolation phase transition and the associated critical temperature.
The Schwinger barrier penetration mechanism modified by Gaussian fluctuations about the average value of the string tension leads to an early (at birth) thermal system \cite{pajares3,bialas}. 
Above the percolation threshold this system is considered to be in the  deconfined phase, which subsequently expands according to Bjorken boost invariant 1D hydrodynamics \cite{bjorken}.


 With an increasing number of strings n there is a progression from isolated individual strings to clusters and then to a large cluster which suddenly spans the area. In two dimensional percolation theory the relevant quantity is the dimensionless percolation density parameter given by \cite{pajares1,pajares2}  
\begin{equation}  
\xi = \frac {N S_{1}}{S_{N}}
\end{equation}
where N is the number of strings formed in the collisions and $S_{1}$
 is  the transverse area of the a single string and $S_{N}$ is the transverse nuclear overlap area. The critical cluster which spans $S_{N}$, appears for
$\xi_{c} \ge$ 1.2 \cite{satz1}. As $\xi$ increases the fraction of $S_{N}$ covered by this spanning cluster increases.

  The color suppression factor F($\xi$), reduces the hadron multiplicity from n$\mu_{0}$ to the interacting string value $\mu$
\begin{equation}
\mu = F(\xi) n\mu_{0}
\end{equation}

\begin{equation}
F(\xi) = \sqrt \frac {1-e^{-\xi}}{\xi}
\end{equation}

To evaluate the initial value of $\xi$ from data, a parameterization of p-p events at 200 GeV  is used to compute the $p_{t}$ distribution \cite{nucleo}

\begin{equation}
dN_{c}/dp_{t}^{2} = a/(p_{0}+p_{t})^{\alpha}
\end{equation}
where a, $p_{0}$, and $\alpha$ are parameters used to fit the data. This parameterization  also can be used for nucleus-nucleus collisions to take into account the interactions of the strings\cite{nucleo}
\begin{equation}
p_{0} \rightarrow p_{0} \left(\frac {\langle nS_{1}/S_{n} \rangle_{Au-Au}}{\langle nS_{1}/S_{n} \rangle_{pp}}\right)^{1/4}
\end{equation}
where $S_{n}$ corresponds to the area occupied by the n overlapping strings.
The thermodynamic limit, i.e. letting n and $S_{n}$ $\rightarrow \infty$  while keeping $\xi$ fixed, is  used to evaluate
\begin{equation}
\langle \frac {nS_{1}}{S_{n}} \rangle = 1/F^{2}(\xi)
\end{equation}
\begin{equation}
dN_{c}/dp_{t}^{2} = \frac {a}{{(p_{0}{\sqrt {F(\xi_{pp})/F(\xi)}}+p_{t})}^{\alpha}}
\end{equation}

In pp collisions  $ \langle nS_{1} / S_{n} \rangle_{pp}$ $\sim$ 1  due to the low string overlap probability. The factor $1-e^{-\xi}$ in Eq. 4 corresponds to the fractional area covered by the spanning cluster. The STAR analysis of charged hadrons for 0-10\% central Au+Au collisions at $\sqrt{s_{NN}}=$200 GeV gives a value $\xi =2.88$ $\pm $ 0.09 \cite{nucleo}.
 

The strong longitudinal chromo-electric fields produce Schwinger-Bialas \cite{pajares3,bialas,swinger1} like radiation with a thermal spectrum, in analogy with the Hawking-Unruh radiation \cite{hawking,unruh,parikh,khar1,khar2,khar3}. Both the Schwinger-Bialas and Hawking-Unruh derivations lead to the same value of the maximum entropy temperature.
Above the critical value of $\xi$, the QGP in CSPM consists of massless constituents (gluons). 
The percolation parameter $\xi$ determines the
cluster size distribution, the temperature T($\xi$) and the transverse momentum in the collision \cite{pajares3}.  
The connection between $\xi$ and the temperature $T(\xi)$ involves the Schwinger mechanism (SM) for particle production.  
In CSPM the Schwinger distribution for massless particles is expressed in terms of $p_{t}^{2}$
\begin{equation}
dn/d{p_{t}^{2}} \sim e^{-\pi p_{t}^{2}/x^{2}}
\end{equation}
with the average string value of $\langle x^{2} \rangle$. Gaussian fluctuations in the string tension ( Bialas) around its mean value
transforms SM into the thermal distribution \cite{bialas}
\begin{equation}
dn/d{p_{t}^{2}} \sim e^{-p_{t}/ \sqrt {2\pi/\langle x^{2} \rangle} }
\end{equation}
with $\langle x^{2} \rangle$ = $\pi \langle p_{t}^{2} \rangle_{1}/F(\xi)$.
The temperature is given by
\begin{equation}
T(\xi) =  {\sqrt {\frac {\langle p_{t}^{2}\rangle_{1}}{ 2 F(\xi)}}}
\end{equation}

In the determination of temperature using Eq.(11) the value of $F(\xi)$ is obtained using the experimental data \cite{nucleo}. We will adopt the point of view that the experimentally determined chemical freeze-out temperature is a good measure of the phase transition temperature, $T_{c}$ \cite{braunmun}.
The single string average transverse momentum  ${\langle p_{t}^{2}\rangle_{1}}$ is calculated at $\xi_{c}$ = 1.2 with the  universal chemical freeze-out temperature of 167.7 $\pm$ 2.6 MeV \cite{bec1}. This gives $ \sqrt {\langle {p_{t}^{2}} \rangle _{1}}$  =  207.2 $\pm$ 3.3 MeV which is close to  $\simeq$200 MeV used previously in the calculation of percolation transition temperature \cite{pajares3}.
Above $\xi_{c}$ =1.2 the size and density of the spanning cluster increases. We use the measured value of $\xi$ = 2.88 to determine the temperature, before the expansion, $T_{i}$  = 193.6$\pm$3.0 MeV of the quark gluon plasma in reasonable agreement with $T_{i}$  = 221$\pm 19^{stat} \pm 19^{sys}$ from the enhanced direct photon experiment measured by the PHENIX Collaboration\cite{phenix}.

 The QGP according to CSPM is born in local thermal equilibrium  because the temperature is determined at the string level. We use CSPM coupled to hydrodynamics to calculate energy density, pressure, entropy density and sound velocity. As mentioned earlier the strings interact strongly to form clusters and produce the pressure at the early stages of the collisions, which is evident from the presence of elliptical flow in CSPM \cite{bautista}. After the initial temperature $ T > T_{c}$ the  CSPM perfect fluid may expand according to Bjorken boost invariant 1D hydrodynamics \cite{bjorken} 

\begin{eqnarray}
\frac {1}{T} \frac {dT}{d\tau} = - C_{s}^{2}/\tau  \\
\frac {dT}{d\tau} = \frac {dT}{d\varepsilon} \frac {d\varepsilon}{d\tau} \\
\frac {d\varepsilon}{d\tau} = -T s/\tau \\
s =(1+C_{s}^{2})\frac{\varepsilon}{T}\\
\frac {dT}{d\varepsilon} s = C_{s}^{2} 
\end{eqnarray}
where $\varepsilon$ is the energy density, s the entropy density, $\tau$ the proper time, and $C_{s}$ the sound velocity.

 Above the critical temperature only massless particles are present in CSPM. 
The initial energy density $\varepsilon_{i}$ above $T_{c}$ is given by \cite{bjorken}
\begin{equation}
\varepsilon_{i}= \frac {3}{2}\frac { {\frac {dN_{c}}{dy}}\langle m_{t}\rangle}{S_{n} \tau_{pro}}
\end{equation}
To evaluate $\varepsilon_{i}$ we use the charged pion multiplicity $dN_{c}/{dy}$ at midrapidity and $S_{n}$ values from STAR for 0-10\% central Au-Au collisions with $\sqrt{s_{NN}}=$200 GeV \cite{levente}. The factor 3/2 in Eq.(17) accounts for the neutral pions. We can calculate $ \langle p_{t}\rangle$ using the CSPM thermal distribution Eqs.(10) and (11). For $0.2 < p_{t} < 1.5 $, $\langle p_{t}\rangle = 0.394 \pm 0.003 $GeV, adding the extra energy required for the rest mass of pions at hadronization $\langle m_{t}\rangle = 0.42 \pm 0.003 $GeV. 
The error on $ \langle p_{t}\rangle$ is due to the error on $T_{i}$.

The dynamics of massless particle production has been studied in QE2 quantum electrodynamics.
QE2 can be scaled from electrodynamics to quantum chromodynamics using the ratio of the coupling constants \cite{wong}. The production time $\tau_{pro}$ for a boson (gluon) is \cite{swinger}  
\begin{equation}
\tau_{pro} = \frac {2.405\hbar}{\langle m_{t}\rangle}
\end{equation}
Using Eqs. (17) and (18) gives  $\varepsilon_{i}$ = 2.27$\pm $0.16 GeV/$fm^{3}$ at $\xi$=2.88. In CSPM the total transverse energy is proportional to $\xi$. 
From the measured value of  $\xi$ and $\varepsilon$ 
it is found  that $\varepsilon$ is proportional to $\xi$ for the range 
$1.2 < \xi < 2.88$, $\varepsilon_{i}= 0.788$ $\xi$ GeV/$fm^{3}$ \cite{nucleo,levente}. 
This relationship has been extrapolated to below  $\xi= 1.2$ and  above $\xi =2.88$  for the energy and entropy density calculations. The Au-Au at $\sqrt{s_{NN}}=$200 GeV data is used to normalize the CSPM $\varepsilon/T^{4}$ values. Figure 1 shows $\varepsilon/T^{4}$ as obtained from CSPM along with the LQCD calculations \cite{bazavov} and the CSPM pressure $3p/T^{4}$.

The number of degrees of freedom (DOF) are related to the energy density 
\begin{equation}
\varepsilon_{i} = \frac {G(T) \pi^{2} T_{i}^{4}}{30 (\hbar c)^{3}}
\end{equation}
At $T_{i}$ the DOF is 37.5$\pm$3.6. At $T_{c}$, $\varepsilon_{c}$=0.95$\pm$0.07 GeV/$fm^{3}$ and 27.7$\pm$2.6 DOF.

\begin{figure}[thbp]
\centering        
\vspace*{-0.2cm}
\includegraphics[width=0.55\textwidth,height=3.0in]{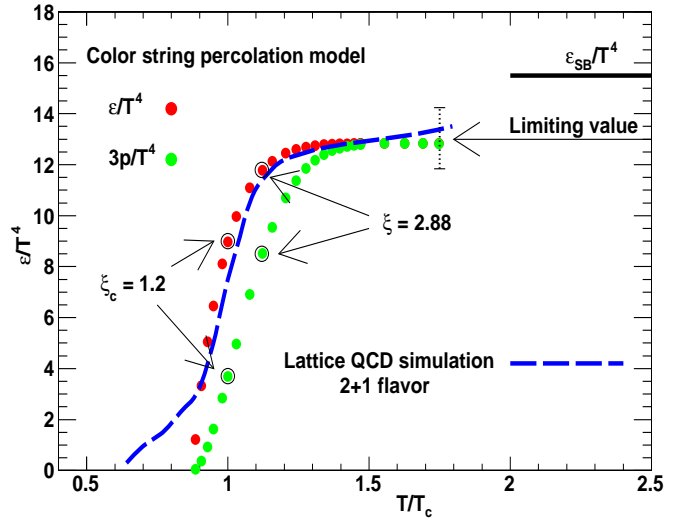}
\vspace*{0.0cm}
\caption{The energy density from CSPM  versus $T/T_{c}^{CSPM}$ (red circles) and Lattice QCD energy density vs  $T/T_{c}^{LQCD}$ (blue dash line) for 2+1 flavor and p4 action \cite{bazavov}. 3p/$T^{4}$ is also shown for CSPM with green circles.} 
\label{perc11}
\end{figure} 

The sound velocity requires the evaluation of s and $ {dT}/{d\varepsilon}$, which can be expressed in terms of $\xi$ and $F(\xi)$. With $q^{1/2}$ = $F(\xi)$ one obtains:

\begin{equation}
\frac {dT}{d\varepsilon} = \frac {dT}{dq}\frac {dq}{d\xi} \frac {d\xi}{d\varepsilon}
\end{equation}
 Then $C_{s}^{2}$ becomes:
\begin{equation}
 C_{s}^{2} = (1+ C_{s}^{2}) (-0.25) \left(\frac {\xi e^{-\xi}}{1- e^{-\xi}}-1\right)
\end{equation}
 for $\xi \geq \xi_{c}$, an analytic function of $\xi$ for the equation of state of the QGP for T $\geq T_{c}$. 

\begin{figure}[thbp]
\centering        
\vspace*{-0.2cm}
\includegraphics[width=0.55\textwidth,height=3.0in]{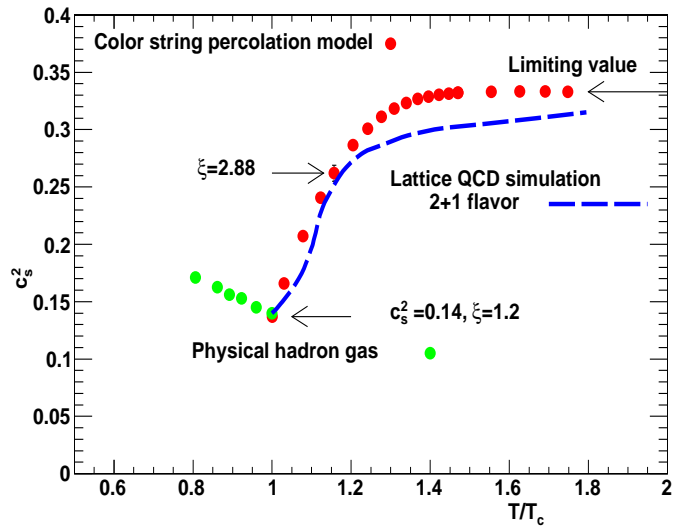}
\vspace*{0.0cm}
\caption{The speed of sound from CSPM  versus $T/T_{c}^{CSPM}$(red circles) and Lattice QCD-p4  speed of sound versus  $T/T_{c}^{LQCD}$(blue dash line)\cite{bazavov}. The physical  hadron gas with resonance mass cut off  M $\leq$ 2.5 GeV is shown as solid green circles \cite{satz}.} 
\label{perc9}
\end{figure}
Figure 2 shows the comparison of $C_{s}^{2}$ from CSPM and LQCD. The LQCD values were obtained using the EOS of 2+1 flavor QCD at finite temperature with physical strange quark mass and almost physical light quark masses \cite{bazavov}. At $T/T_{c}$=1 the CSPM 
 and LQCD agree with the $C_{s}^{2}$ value of the physical hadron gas with resonance mass truncation M $\leq$ 2.5 GeV \cite{satz}.

The entropy density $s/T^{3}$ is computed using Eq. (15) as shown in Fig.3 along with the LQCD results. CSPM is in excellent agreement with the LQCD calculations in the phase transition region for $T/T_{c} \leq $1.5. 
It is noteworthy that CSPM at $\xi > \xi_{c}$ exhibits gluon saturation effects similar to the Color Glass Condensate (CGC) \cite{larry}. The saturation scale $Q_{s}$ in CGC corresponds to $ {\langle p_{t}^{2} \rangle_{1}}/F(\xi)$ in CSPM.

\begin{figure}[thbp]
\centering        
\vspace*{-0.2cm}
\includegraphics[width=0.55\textwidth,height=3.0in]{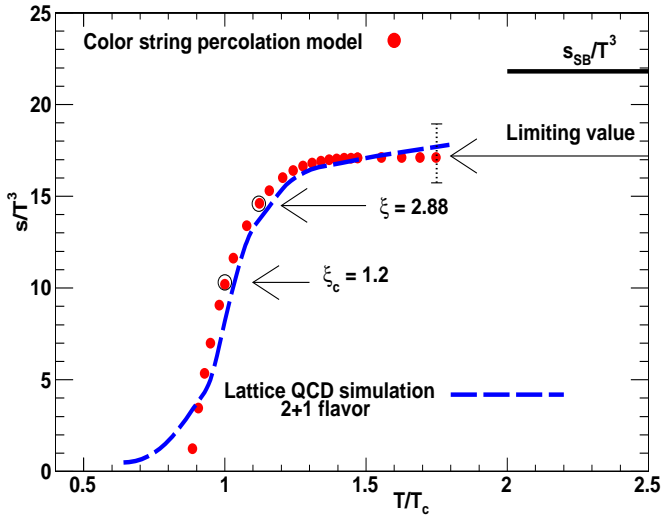}
\vspace*{0.0cm}
\caption{The entropy density from CSPM  versus $T/T_{c}^{CSPM}$(red circle) and Lattice QCD entropy density vs  $T/T_{c}^{LQCD}$(blue dash line) \cite{bazavov}.} 
\label{perc10}
\end{figure}
In summary, the present analysis consists of using the $\xi$ values from data, identifying the universal chemical temperature with the critical percolation density parameter $\xi_{c}$, establishing the direct proportionality between $\varepsilon_{i}$ and $\xi$ and using the functional dependence of $T(\xi)$ and $\varepsilon_{i}$ to obtain $C_{s}^{2}$. 
Thus we can  determine the temperature dependence of energy density, pressure and entropy density in the color string percolation picture.     
The results are also in agreement with lattice QCD in the phase transition region, when the results are plotted with respect to $T/T_{c}^{CSPM}$ and  $T/T_{c}^{LQCD}$. The value of $C_{s}^{2}$=0.14 is in agreement with the physical resonance gas value at the critical temperature. The non-interacting high temperature limit for  $C_{s}^{2}$ = 0.33 is reached at  $T \sim 1.5T_{c}$.  

The percolation critical transition is known to represent a continuous phase transition. In central Au-Au collisions at $\sqrt{s_{NN}}=$ 200 GeV the QCD to hadron phase transition for baryon density $\mu_{B} \sim $0 is believed to be a cross-over transition which does not have a latent heat \cite{fodor}. The CSPM EOS correctly describes the QCD to hadron cross-over transition and provides an answer to the question of the origin of universal temperature observed in A-A, p-p and $e^{+} e^{-}$ collisions.

The percolation analysis of the color sources applied to STAR data at RHIC provides a compelling argument that the QGP is formed in central Au-Au collisions at $\sqrt{s_{NN}}=$ 200 GeV. It also suggests that the QGP is produced in all soft high energy high multiplicity collisions when the string density exceeds the percolation transition. A further definitive test of CSPM can be made at LHC energies by comparing  hadron-hadron and nucleus-nucleus collisions.

We express our thanks to C. Pajares and N. Armesto for many fruitful discussions. This research was supported by the Office of Nuclear Physics within the U.S. Department of Energy  Office of Science under Grant No. DE-FG02-88ER40412.



\begin{thebibliography}{}

\bibitem{bec1}
F. Becattini, P. Castorina, A. Milov, H. Satz, Eur. Phys. J. C{\bf 66}, 377 (2010). 
\bibitem{bec2}
F. Becattini, P. Castorina, J. Manninen, H. Satz, Eur. Phys. J. C{\bf 56}, 493 (2008).
\bibitem{satz2}
H. Satz, Eur. Phys. J. Special Topics {\bf 155}, 167 (2008).
\bibitem{stat1}
A. Andronic et al, Phys. Lett. B{\bf 675}, 312 (2009).
\bibitem{bazavov}
A. Bazavov et al., Phys. Rev. D{\bf 80}, 014504 (2009).
\bibitem{pajares1}
M. A. Braun, C. Pajares, Eu. Phys. J. C{\bf 16}, 349 (2000).
\bibitem{pajares2}
M. A. Braun, F. del Moral, C. Pajares, Phys. Rev. C {\bf 65}, 024907 (2002).
\bibitem{diasde} 
J. Dias de Deus, E. G. Ferreiro, C. Pajares, R. Ugoccioni, Eur. Phys. J. C{\bf 40}, 229 (2005).
\bibitem{bautista} 
I. Bautista, L. Cunqueiro, J. Dias de Deus, C. Pajares, J. Phys. G{\bf 37}, 015103 (2010).
\bibitem{pajares3} 
 J. Dias de Deus, C. Pajares, Phys. Lett.  B{\bf 642}, 455 (2006). 
\bibitem{LRC}
B. I. Abelev et al., (STAR Collaboration), Phys. Rev. Lett. {\bf 103}, 172301 (2009).
\bibitem{bialas}
A. Bialas, Phys. Lett. B{\bf 466}, 301 (1999).
\bibitem{bjorken}
J. D. Bjorken, Phys. Rev. D{\bf 27}, 140 (1983).
\bibitem{satz1}
 H. Satz, Rep. Prog. Phys. {\bf 63}, 1511 (2000).
\bibitem{nucleo}
B. K. Srivastava, R. P. Scharenberg, T. Tarnowsky, (STAR Collaboration), Nukleonika {\bf 51}, s109 (2006).
\bibitem{swinger1} 
J. Schwinger, Phys. Rev. {\bf 82}, 664 (1951).
\bibitem{hawking}
S. W. Hawking, Commun. Math. Phys. {\bf 43}, 199 (1975).
\bibitem{unruh}
W. G. Unruh, Phys. Rev. D{\bf 14}, 870 (1976).
\bibitem{parikh}
M. K. Parikh, F. Wilczek, Phys. Rev. Lett. {\bf 85}, 5042 (2000).
\bibitem{khar1}
D. Kharzeev, K. Tuchin, Nucl. Phys. A{\bf 753}, 316 (2005).
\bibitem{khar2}
D. Kharzeev, E. Levin, K. Tuchin, Phys. Rev. C{\bf 75}, 044903 (2007).
\bibitem{khar3}
P. Castorina, D. Kharzeev, H. Satz, Eur. Phys. J. C{\bf 52}, 187 (2007).
\bibitem{braunmun}
P. Braun-Munzinger, J. Stachel, Christof Wetterich, Phys. Lett. B{\bf 596}, 61 (2004).
\bibitem{phenix}
A. Adare et al., (PHENIX Collaboration), Phys. Rev. Lett. {\bf 104}, 132301 (2010). 
\bibitem{levente}
B. I. Abelev et al., (STAR Collaboration), Phys. Rev. C{\bf 79}, 34909 (2009). 
\bibitem{wong}
C. Y. Wong, Introduction to high energy heavy ion collisions (World Scientific,1994).
\bibitem{swinger} 
J. Schwinger, Phys. Rev. {\bf 128}, 2425 (1962).
\bibitem{satz}
 P. Castorina, J. Cleymans, D. E. Miller, H. Satz, arXiv:hep-ph/0906.2289v1. 
\bibitem{larry} 
J. Schaffner-Bielich, D. Kharzeev, L. McLerran, R. Venugopalan, Nucl. Phys. A{\bf 705}, 494 (2002).
\bibitem{fodor}
Z. Fodor, PoSLATT2007:011 (2007).
\end{thebibliography}
\end{document}